\documentclass[a4paper,twocolumn,showpacs,,amsmath,amssymb,pra,superscriptaddress]{revtex4-1}

\usepackage[draft]{hyperref}
\usepackage{graphicx}
\usepackage{dcolumn}
\usepackage{amssymb,amsmath}
\usepackage{color}

\newcommand{\ba}{\begin{array}}
\newcommand{\ea}{\end{array}}
\newcommand{\be}{\begin{equation}}
\newcommand{\ee}{\end{equation}}
\newcommand{\bea}{\begin{eqnarray}}
\newcommand{\eea}{\end{eqnarray}}

\newcommand{\bfE}{\textbf{E}}
\newcommand{\bfc}{\textbf{c}}
\newcommand{\bfx}{\textbf{x}}
\newcommand{\bfy}{\textbf{y}}
\newcommand{\bfz}{\textbf{z}}

\begin{document}

\title{Ultra-broadband gradient-pitch Bragg-Berry mirrors}

\author{Mushegh Rafayelyan}
\address{Univ. Bordeaux, CNRS, LOMA, UMR 5798, F-33400 Talence, France}

\author{Gonzague Agez}
\address{CEMES, CNRS, University Paul-Sabatier, F-31055 Toulouse, France}

\author{Etienne Brasselet}
\email{etienne.brasselet@u-bordeaux.fr}
\address{Univ. Bordeaux, CNRS, LOMA, UMR 5798, F-33400 Talence, France}

\date{\today}

\begin{abstract}
The realization of geometric phase optical device operating over a broad spectral range is usually confronted with intrinsic limitations depending of the physical process at play. Here we propose to use chiral nematic liquid crystal slabs with helical ordering that varies in three dimensions. Namely, gradient-pitch cholesterics endowed with in-plane space-variant angular positioning of the supramolecular helix. By doing so, we show that the recently introduced Bragg-Berry mirrors [Opt. Lett. {\bf 41}, 3972-3975 (2016)] can be endowed with ultra-broadband spectral range. Experimental demonstration is made in the case of ultra-broadband optical vortex generation in the visible domain. These results offer practical solution to the polychromatic management of the orbital angular momentum of light combining the circular Bragg reflection of chiral media with the Berry phase.
\end{abstract}




\maketitle

\section{Introduction}
\label{section:intro}

\subsection{Context}

Beam shaping is a basic need for photonic technologies (e.g., optical imaging, optical information and communication, optical manipulation, laser processing of materials) and nowadays there are a lot of techniques to control at will the spatial distribution of intensity, phase and polarization. In general this involves the use of optical elements that are designed to operate efficiently for a given wavelength while user-friendly devices or some applications preferably require large spectral bandwidth.

This may become challenging when dealing with optical phase structuring since the latter may be strongly dependent on wavelength. Here we deal with the elaboration of broadband specific optical elements. Namely, anisotropic and inhomogeneous optical elements relying on so-called geometric phase, see for instance the early proposal of a geometric phase lens by Bhandari \cite{bhandari_physrep_1997}. Such elements have great potential for polychromatic arbitrary phase shaping once combined with material structuring add-on features, as discussed hereafter.

Let us consider an ideal lossless geometric phase optical element as a transparent dielectric half-wave plate characterized by a transverse two-dimensional distribution of the optical axis orientation angle, $\psi_{\rm 2D}(x,y)$. Such a device imparts to an incident circularly polarized plane wave propagating along the $z$ axis a space-variant phase $\pm2\psi_{\rm 2D}(x,y)$ that is of a geometric nature, where the $\pm$ sign depends on the incident polarization handedness. The underlying physics is one of the numerous manifestations of the spin-orbit interactions of light \cite{bliokh_np_2015}. Noticeably, such a behavior is wavelength-dependent since the half-wave retardation condition is satisfied for a specific wavelength. To solve such an issue, metamaterials offer an elegant option and allow considering the realization of broadband geometric phase flat optical elements. However, the realization of lossless scalable geometric phase metasurfaces operating in the visible domain remains technologically difficult. In practice, promising  developments came from patterned liquid crystals technologies where the possibility to achieve arbitrary two-dimensional orientational patterns $\psi_{\rm 2D}(x,y)$ \cite{kim_optica_2015, chen_pr_2015} can be endowed with an additional degree of freedom along the third dimension, namely, $\psi_{\rm 3D}(x,y,z)$.

The principle of the latter three-dimensional approach finds its roots in the early contributions by Destriau and Prouteau \cite{destriau_jpr_1949} and Pancharatnam \cite{pancharatnam_pias_1955_part1, pancharatnam_pias_1955_part2} who proposed to use a discrete set of uniform waveplates (respectively, two and three) oriented along specific direction in order to achieve achromatic waveplates. Nowadays, space-variant multi-twists waveplates can be realized and various liquid crystal geometric phase optical elements having enhanced achromatic performances \cite{komanduri_oe_2013} have been demonstrated experimentally, for instance geometric phase prisms, lenses or vortex generators \cite{oh_ol_2008, li_spie_2012, tabiryan_oe_2016}. However, strictly speaking, the half-wave retardance condition is not realized but its near fulfillment is achieved over a broad spectral bandwidth.

\begin{figure}[t!]
\centering\includegraphics[width=1\columnwidth] {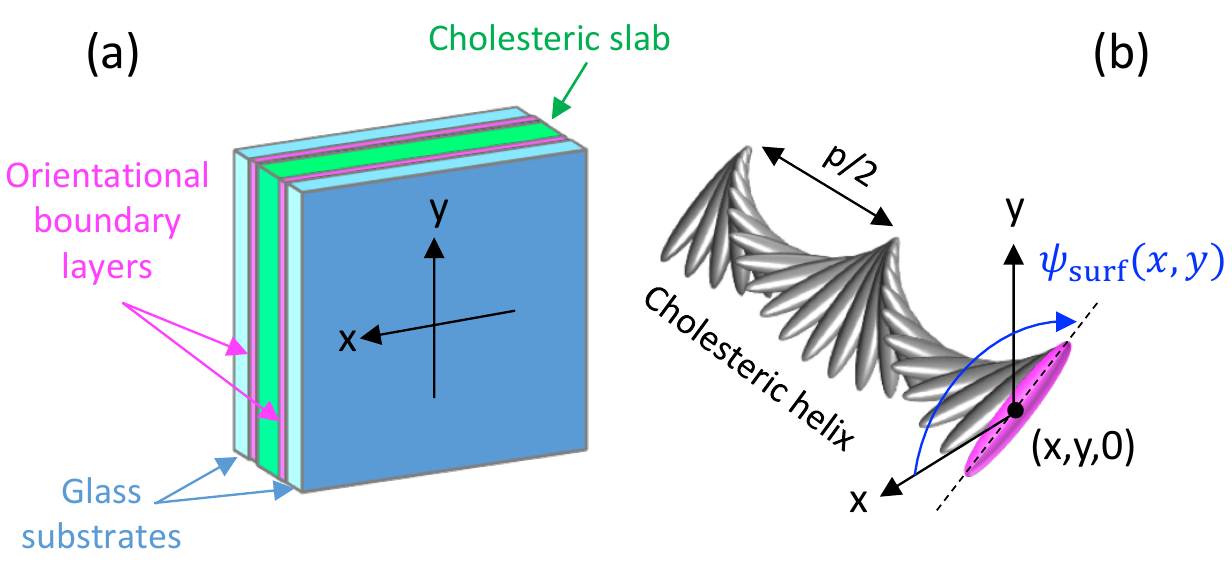}
\caption{
\label{fig:BBOE}
(a) Sketch of a typical Bragg-Berry optical element. (b) Representation of the supramolecular cholesteric helix with constant pitch $p_0$ and handedness $\chi = +1$, which is associated with orientational surface boundary condition $\psi_{\rm surf} (x,y)$.}
\end{figure}

Recently, an alternative approach has been explored based on the use of chiral nematic liquid crystals (also called cholesterics), which are known to satisfy the half-wave retardance condition in reflection over large spectral bandwidth. The idea consists to combine helical structuring of the optical axis of cholesterics along the propagation direction with space-variant optical axis orientation in the transverse plane. This can be described by
\be
\label{eq:psi3D_p0_general}
\psi_{\rm 3D}(x,y,z) = 2\pi\chi z/p_0 + \psi_{\rm surf} (x,y)\,,
\ee
where $\chi=\pm1$ refers to the handedness of the supramolecular helix and the pitch $p_0$ corresponds to the distance over which the director rotates by $2\pi$ around the $z$ axis. In addition, $\psi_{\rm surf} (x,y)$ refers the two-dimensional orientational boundary condition at $z=0$ and $z=L$, as shown in Fig.~\ref{fig:BBOE}.

To understand how such space-variant chiral anisotropic media associated with three-dimensional pattern may impart to an incident circularly polarized plane wave propagating along the $z$ axis an optical phase spatial distribution of the form $\pm2\psi_{\rm surf}(x,y)$, we first recall hereafter the basics of the optics of uniform cholesterics.

\subsection{Background}
\label{subsection:background}

\begin{figure}[b!]
\centering\includegraphics[width=1\columnwidth] {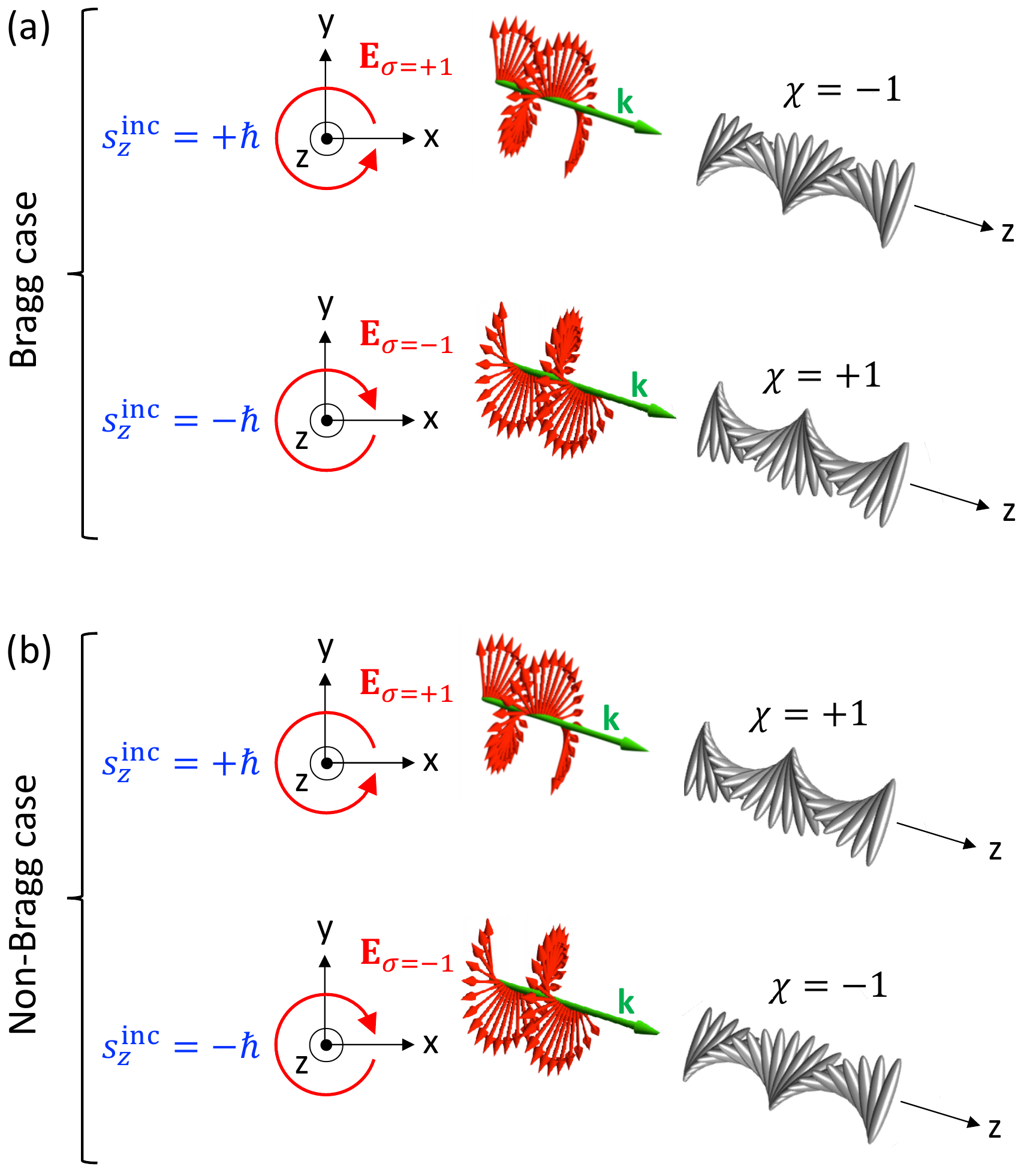}
\caption{
\label{fig:bragg-nobragg}
Summary of various features associated the electric field of circularly polarized light and the helical molecular orientation of a cholesteric for $\sigma=\pm1$ and $\chi = \pm1$. (a) Bragg case. (b) Non-Bragg case. From left to right: incident optical spin angular momentum projection along the propagation direction; dynamics of the tip of the electric field vector in the transverse plane; helical structuring of the electric field vector at fixed time; supramolecular cholesteric helix.}
\end{figure}

Uniform cholesterics are characterized by $\psi_{\rm surf}(x,y)=\psi_0$ with $\psi_0$ constant. They are well-known for their ability to reflect selectively one of the two circular polarization states for wavelengths belonging to the photonic bandgap defined by $p n_\perp < \lambda < p n_\parallel$, where $n_{\parallel,\perp}$ are respectively the refractive indices along and perpendicular to the liquid crystal director \cite{belyakov_book_92, oswald_book_nematic}. This phenomenon is called the circular Bragg reflection \cite{faryad_aop_2014}. More precisely, let us consider a circularly polarized plane wave propagating towards $+z$ that we described by a complex electric field $\bfE_\sigma = E_0 \exp(-i\omega t + k_0 z) \bfc_\sigma$ where $E_0$ is a constant, $\omega$ is the angular frequency, $t$ is the time, $k_0$ is the free-space wave vector, and $\bfc_\sigma = (\bfx + i\sigma \bfy)/\sqrt{2}$ is the circular polarization unit basis in the unit Cartesian coordinate system $(\bfx, \bfy, \bfz)$ with $\sigma=\pm1$ being the helicity of light. When a uniform cholesteric slab is thick enough with respect to the pitch and wavelengths inside the bandgap, a circularly polarized light propagating along the cholesteric helix axis is almost fully reflected if the helix formed by the tip of its electric field vector has the same handedness as the cholesteric helix \cite{remark_almost}, i.e., $\sigma = -\chi$. On the other hand, light is almost fully transmitted when the optical and material helices have opposite handednesses \cite{remark_almost}, i.e., $\sigma = +\chi$. We refer to Fig.~\ref{fig:bragg-nobragg} for an illustrated summary of the relative helical structuring of the electric field and the cholesteric molecular orientation in the Bragg [panel (a)] and non-Bragg [panel (b)] situations.\\

\begin{figure*}[t!]
\centering\includegraphics[width=2.08\columnwidth] {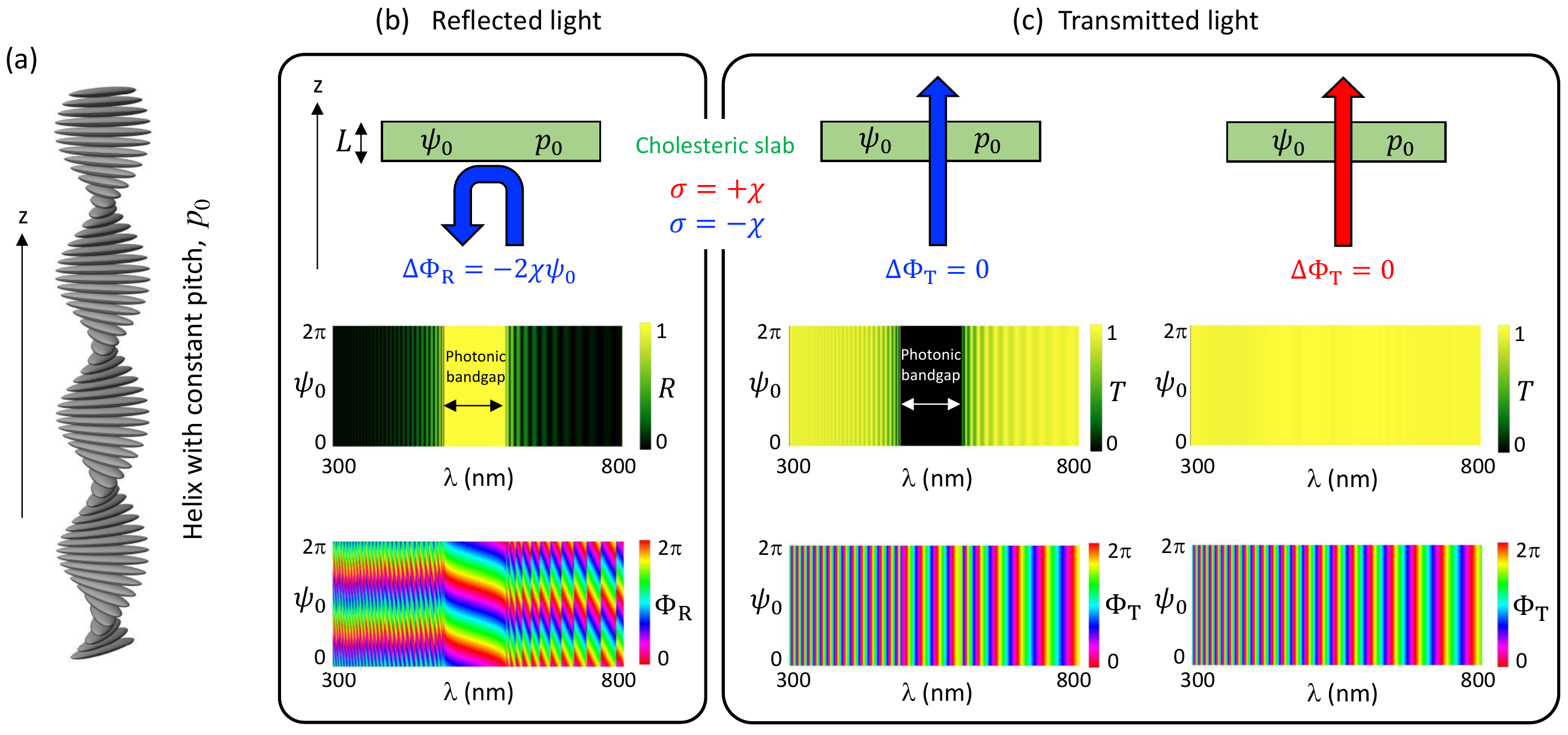}
\caption{
\label{fig:BBM}
(a) Illustration of the supramolecular helix with constant pitch $p_0$ and handedness $\chi=+1$. (b) Calculated reflectance ($R$) and phase ($\Phi_R$) spectra of the helicity-preserved Bragg-reflected light in the case of a uniform cholesteric slab having a constant pitch, for all possible orientational boundary conditions $0 \leq \psi_0 \leq 2\pi$. (c) Same as in box (b) for the transmittance ($T$) and phase ($\Phi_T$) of the transmitted light, for which the two cases $\sigma=\pm \chi$ must be considered separately. Simulations are made using $4 \times 4$ Berreman formalism \cite{berreman_josa_72} with $n_{\rm glass}=1.52$, $L=7~\mu$m, $n_\parallel=1.71$, $n_\perp =1.43$, $p=350~$nm, $\chi=+1$ and considering normally incident plane wave.}
\end{figure*}

Importantly, Bragg reflection preserves the electric field helix handedness, hence flips the spin angular momentum projection along the $z$ axis per photon from $s_z = -\chi \hbar$ (incident light) to $s_z = +\chi \hbar$ (Bragg-reflected light). However, for wavelengths outside the bandgap, only a fraction of the incident light is Bragg-reflected while the rest is transmitted without spin-flip. Finally, light is almost fully transmitted for $\sigma=+\chi$ without spin-flip \cite{remark_almost}. The situation is quantitatively summarized in Fig.~\ref{fig:BBM} where blue and red arrows respectively refer to helicity $\sigma=\mp\chi$ for the propagating light. In addition, for the considered cases, we also show numerical simulations of the reflectance ($R$) and transmittance ($T$) spectra, which are demonstrated to be independent on $\psi_0$.

The figures \ref{fig:BBM}(b) and \ref{fig:BBM}(c) also display the dependence of the reflected ($\Phi_R$) and transmitted ($\Phi_T$) phase spectra on $\psi_0$, which are evaluated according to $\Phi_{R,T} = \arg(E_{R,T})$, where $E_{R,T}$ are the complex reflected and transmitted electric field amplitudes at $z=0$ and $z=L$, respectively. These simulations unveil qualitatively \cite{remark_qualitative} the possibility to shape the transverse distribution of the optical phase of the Bragg-reflected light. Indeed, arbitrary phase difference between two points in the $(x,y)$ plane can be provided by using space-variant orientational boundary conditions noting that $\Delta \Phi_R = \Phi_R(\psi_0) - \Phi_R(0) = -2\chi \psi_0$, see Fig.~\ref{fig:BBM}(b). That is to say, a space-variant cholesteric slab characterized by $\psi_{\rm surf}(x,y)$ according to Eq.~(\ref{eq:psi3D_p0_general}) is expected to act as a reflective spatial light modulator imparting a phase spatial distribution of the form $\pm2\psi_{\rm surf}(x,y)$ to an incident circularly polarized light field with helicity $\sigma = -\chi$. In contrast, $\Delta \Phi_T = 0$ for $\sigma = \pm\chi$ and therefore no transverse phase shaping is anticipated in transmission. Such a behavior results from the fact that the  transmitted phase does not depend on $\psi_0$ although it is of course strongly dependent on wavelength, see Fig.~\ref{fig:BBM}(c).

The optical properties of above discussed space-variant cholesterics with constant pitch have been experimentally demonstrated recently, with already various demonstrations in the framework of prisms and lenses \cite{kobashi_np_2016}, and optical vortex generators \cite{rafayelyan_prl_2016, rafayelyan_ol_2016, kobashi_prl_2016}. These recent works thus unveiled a novel class of geometric phase reflective optical elements endowed with intrinsic polychromatic features.

\subsection{`Bragg-Berry mirror' terminology}
\label{subsection:terminology}

Physically, the discussed geometric phase reflective optical elements rely on the combination of two distinct optical effects that respectively confer unique properties to the fabricated optical devices. Namely, polychromatic response and helicity-controlled phase structuring. Indeed, the latter  characteristics respectively rely on (i) circular Bragg reflection of chiral anisotropic media \cite{faryad_aop_2014} and (ii) geometric (Berry) phase associated with the coupling between intrinsic optical spin angular momentum and rotations of coordinates \cite{bliokh_np_2015}. This invited us to suggest the `Bragg-Berry mirror' terminology in a previous work \cite{rafayelyan_ol_2016} in order to emphasize these underlying key features.

However, it should be clear that such Bragg-Berry planar mirrors are not standard flat reflectors since light preserves its helicity upon reflection and may acquire arbitrary space-variant phase transverse spatial distribution. Also, we note that the geometric phase nature of such a beam shaping has been experimentally identified in Ref.~\citenum{rafayelyan_prl_2016} with two independent experiments that underpin the universality of the geometric phase regarding temporal and spatial continuous rotations, as discussed in Ref.~\citenum{bliokh_prl_2008}. In addition, an experiment based on discrete jump of orientation by an angle of $\pi/2$ leading to a $\pi$ phase shift has been reported in Ref.~\citenum{barboza_prl_2016}.

\begin{figure*}[t!]
\centering\includegraphics[width=2.08\columnwidth] {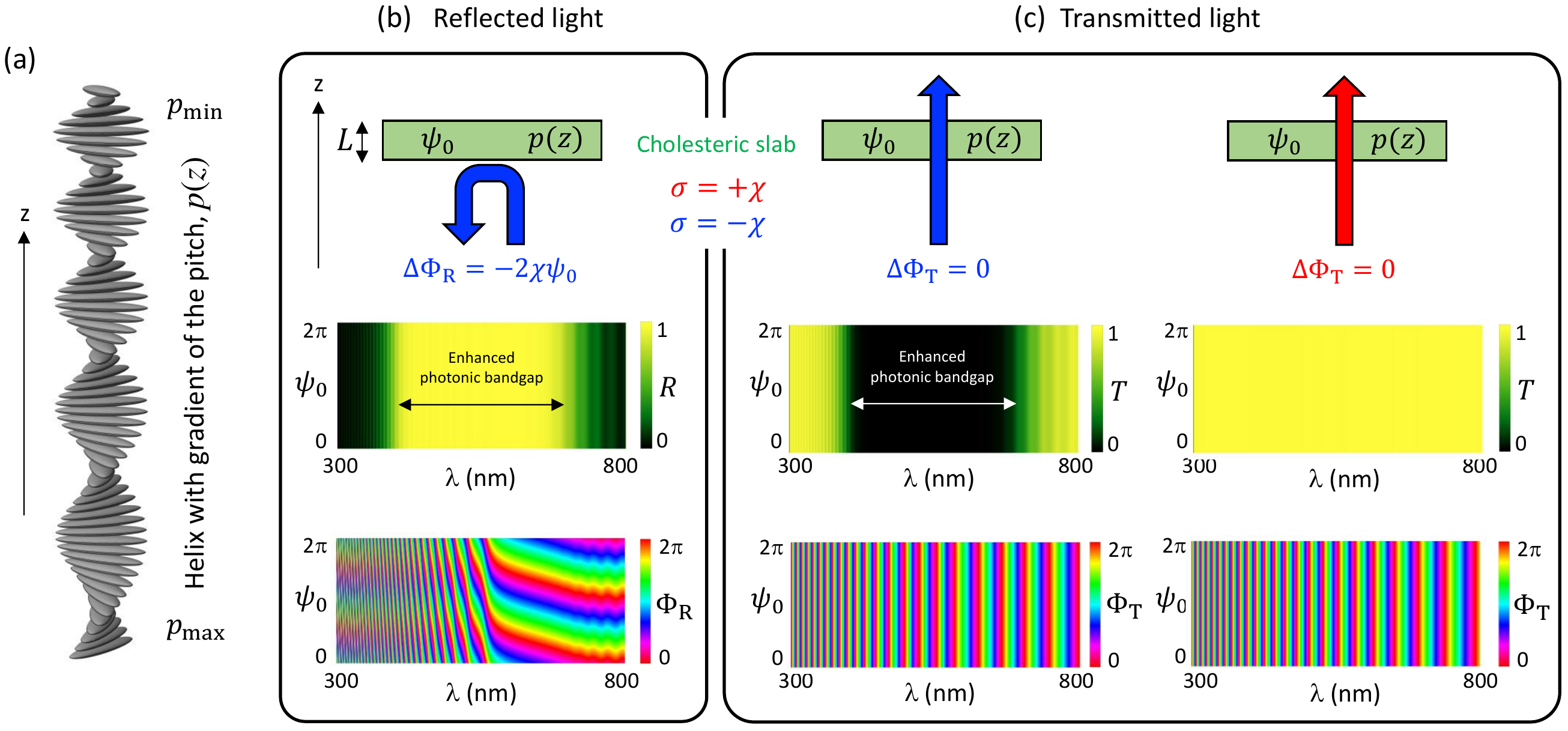}
\caption{
\label{fig:BBMgrad}
Same as Fig.~\ref{fig:BBM} but for a uniform cholesteric with a gradient of the pitch described by $p(z) = p_{\rm max} - (p_{\rm max} - p_{\rm min})z/L$ with $p_{\rm min} = 340$~nm and $p_{\rm max} = 430$~nm.}
\end{figure*}

\subsection{Position of the problem}
\label{subsection:position}

As said in section \ref{subsection:background}, uniform cholesterics with constant pitch have only first-order circular photonic bandgap defined by the wavelength range $n_\perp p_0 < \lambda < n_\parallel p_0$, which is typically less than 100~nm in the visible domain (liquid crystals have usually optical anisotropy up to $n_\parallel - n_\perp \sim 0.3$). However, it is known that  cholesterics having non-uniform pitch may exhibit higher-order photonic bandgaps or broaden their photonic bandgap \cite{broer_am_1999}. In particular, cholesteric helix characterized by a pitch gradient allows to achieve circular Bragg reflection over the full visible domain, which can be realized using photo-induced \cite{broer_nature_1995} or thermally induced \cite{mitov_epjb_1999} as experimentally demonstrated in earlier works. For recent in-depth overviews we refer to review papers \cite{mitov_advmat_2012, balamurugan_rfp_2016}. The back-home message here is that the photonic bandgap can be extended from $n_\perp p_0 < \lambda < n_\parallel p_0$ in the case of constant pitch to $n_\perp p_{\rm min} < \lambda < n_\parallel p_{\rm max}$ estimated in the case of gradient-pitch cholesterics defined by a $z$-dependent pitch $p(z) = p_{\rm max} - (p_{\rm max} -p_{\rm min} )z/L$ \cite{broer_am_1999}.

In practice, the gradient-pitch bandwidth can be several times larger that the constant-pitch bandwidth. Accordingly, we propose to combine the previously demonstrated topological beam shaping capabilities of constant-pitch Bragg-Berry mirrors with the ultra-broadband performances of gradient-pitch cholesterics. This would provide with reflective geometric phase optical elements operating efficiently over the full visible spectrum. Of course, such a principle could be extended to larger wavelength regions by appropriate setting of pitch values.

Our work is organized such as, firstly, we address the case of uniform gradient-pitch Bragg mirrors in section \ref{section:uniform}, where the existence of geometric phase in the Bragg-reflected light is explored numerically and demonstrated experimentally. Then, without loss of generality, we address in section \ref{section:nonuniform} the experimental realization of ultra-broadband gradient-pitch Bragg-Berry mirrors in the context of optical vortex generation. Finally, we summarize the study and discuss a few prospective issues.

\section{Uniform gradient-pitch Bragg mirrors}
\label{section:uniform}

\subsection{Geometric phase of reflected field: numerics}
\label{subsection:uniform-numerics}

Following the numerical approach presented in section \ref{subsection:background} that allows to anticipate the existence of a geometric phase for the Bragg-reflected light in the case of uniform cholesterics with constant pitch [see Fig.~\ref{fig:BBM}(b)], we perform a similar set of simulations in the case of uniform gradient-pitch cholesterics described by
\be
\label{eq:psi3D_p(z)_uniform}
\psi_{\rm 3D}(z) = 2\pi\chi z/p(z) + \psi_0\,,
\ee
with the space-variant pitch $p(z)$ given by
\be
\label{eq:p(z)}
p(z) = p_{\rm max} - (p_{\rm max} -p_{\rm min} )z/L\,,
\ee
where the values of $p_{\rm min}$ and $p_{\rm max}$ are chosen to have identical central wavelength for the photonic bandgap, which eases the comparison between the two situations. The results are presented in Fig.~\ref{fig:BBMgrad} that is organized in a similar fashion as in Fig.~\ref{fig:BBM}.

As shown in Fig.~\ref{fig:BBMgrad}(b) and \ref{fig:BBMgrad}(c), the expected substantial increase of the photonic bandgap bandwidth for $\sigma=-\chi$ is observed. Noteworthy, regarding the reflected phase spectrum for $\sigma=-\chi$ and the transmitted phase spectra for $\sigma = \pm\chi$, the conclusions are unchanged with respect to the case of constant pitch. Namely, a non-uniform pitch neither prevents the existence of the geometric phase nor its expression $\Delta \Phi_R = -2\chi \psi_0$, while significantly widens the photonic bandgap. Therefore, this underpins our proposition presented in section \ref{subsection:position} and invites us to implement its experimental validation, which is performed in what follows.

\subsection{Sample  preparation and characterization}
\label{subsection:uniform-preparation}

\begin{figure}[b!]
\centering\includegraphics[width=1\columnwidth]{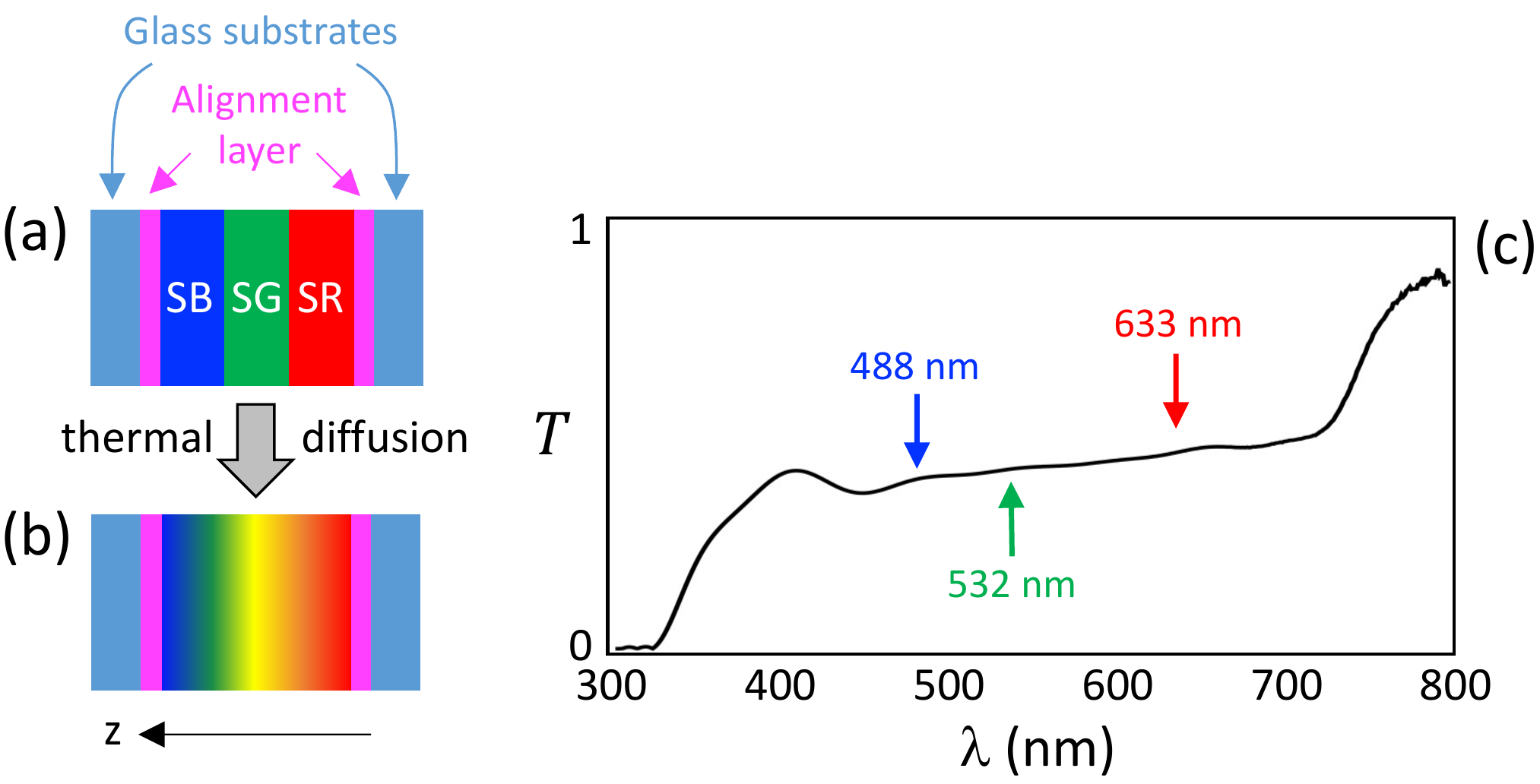}
\caption{
\label{fig:sample}
Sketch of the initial three-layer stack of cholesteric oligomer (SB, SG, SR) sandwiched between tow glass substrates coated with surface-alignment layers providing the orientational boundary conditions before (a) and after (b) thermal treatment that leads to promote the inter-diffusion of the three layers. (c) Typical transmission spectrum of a fabricated uniform gradient-pitch cholesteric sample under unpolarized illumination.}
\end{figure}

We used cholesteric oligomers from Wacker Chemie GmbH with $n_\parallel = 1.72$ and $n_\perp = 1.42$ effective extraordinary and ordinary refractive indices. The molecular structure of the material consists of a siloxane cyclic chain to which is attached, via aliphatic spacers, two types of side chains. Namely, an achiral mesogen and a chiral cholesterol-bearing mesogen. The pitch of the helical structure and therefore the reflection wavelength depends on the molar percentage of the chiral mesogene in the molecule. This percentage is 31\% in the case of \textit{Silicon Red} compound (SR), 40\% for \textit{Silicon Green} compound (SG) and 50\% for \textit{Silicon Blue} compound (SB). The cholesteric phase appears between 40--50$^\circ$C, which corresponds to the glass-transition temperature range, and 180--210$^\circ$C, which corresponds to the clearing temperature range. By freezing the film in a glassy solid state, the cholesteric structure and its optical properties are kept at room temperature in a perennial manner.

A three-layer sample was elaborated by stacking three different $10~\mu$m-thick cholesteric oligomers layers, namely, SB/SG/SR, as shown in Fig.~\ref{fig:sample}(a), according to the protocol described hereafter. The SR and SB films were prepared by confinement between a glass substrate covered by polyimide surface anchoring layers and a microscope coverslip and the SG layer was confined between two microscope coverslips. The samples were kept at $80^\circ$C (cholesteric phase) for 10 minutes to form textures with the cholesteric helix being perpendicular to the substrates that follow either the orientational boundary conditions provided by uniform surface anchoring layer or the planar degenerate orientation provided by the microscope coverslips. Then the samples are placed in the freezer for 15 minutes $-18^\circ$C in order to obtain glassy solid layers. At this temperature the coverslips can be easily taken off and obtained free-standing SG layer was sandwiched between the SR and SB semi-free-standing films. Finally, the SB/SG/SR three-layer sample was annealed several times at 80$^\circ$C during 5 minutes to promote the inter-diffusion of the three compounds and obtain a constant gradient of the cholesteric pitch along $z$.

The resulting ultra-broadband circular Bragg reflection is illustrated in Fig.~\ref{fig:sample}(c) that shows the typical transmission spectrum of a uniform gradient-pitch cholesteric sample. Indeed, the photonic bandgap covers almost all the visible domain. Still, we note that the linear gradient-pitch description given by Eq.~(\ref{eq:p(z)}) is an idealization. Indeed, the actual pitch profile along the $z$ axis evolves during the diffusion process \cite{gevorgyan_jnano_2015}. Therefore, in our case, it depends on the time at which the cholesteric has been quenched in the glassy state.

\subsection{Dynamical geometric phase experiment}
\label{subsection:uniform-experiment}

\begin{figure}[t!]
\centering\includegraphics[width=1\columnwidth]{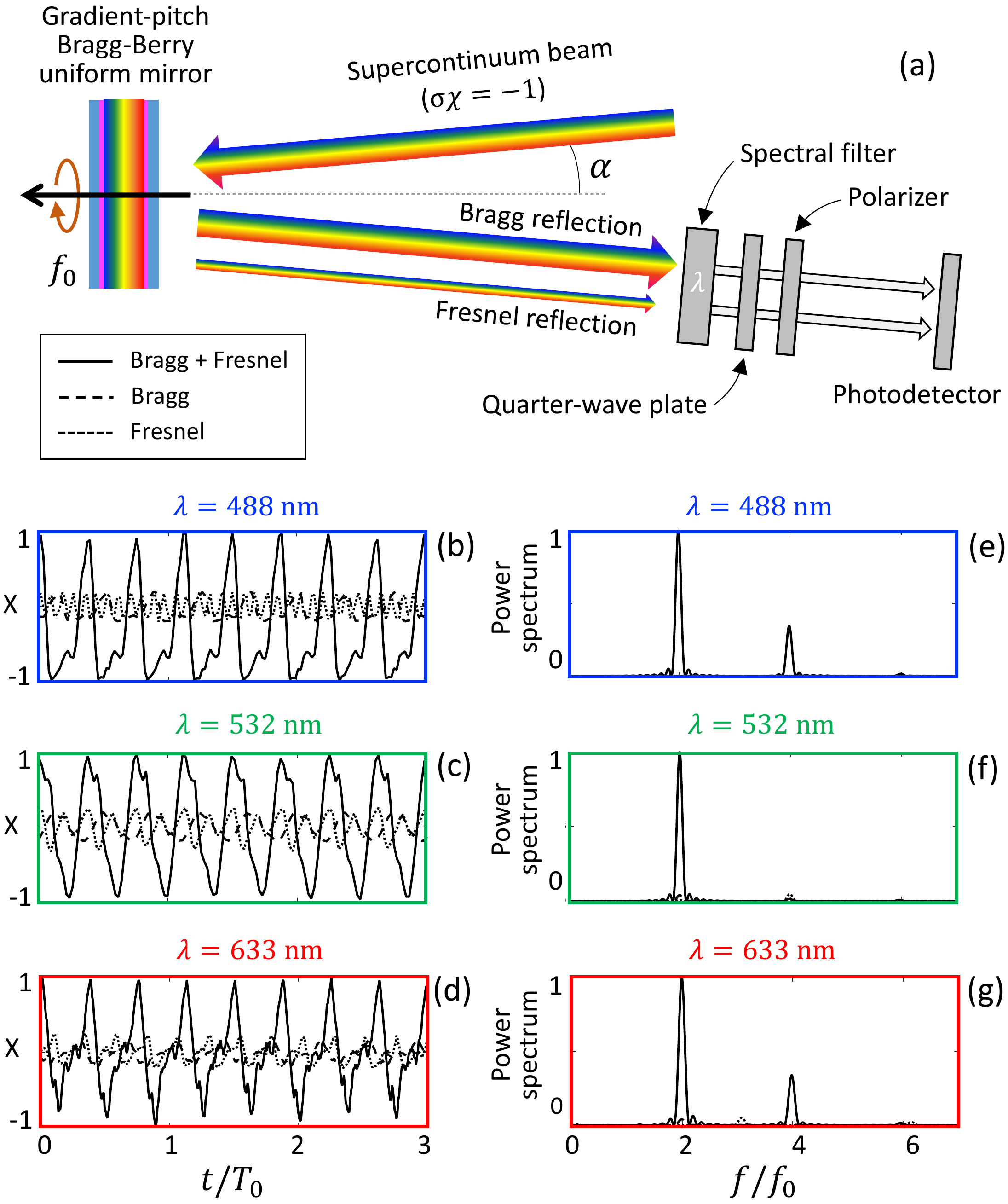}
\caption{
\label{fig:dynamic}
(a) Dynamical geometric phase experimental set-up. Slightly oblique incidence (the angle $\alpha$ equals a few degrees) allows to readily analyze the reflected light. Interference spectral filters for $\lambda = 488$, 532 and 633~nm and having a 3 nm full-width half-maximum transmission spectrum have been used. Thin/thick arrows that  respectively depict the Fresnel/Bragg reflected light beam refers to incident light reflected at air/glass interface and to circular Bragg reflection from the cholesteric. Sample is rotated at frequency $f_0 \simeq 0.05$~Hz. (b,c,d) $X = [s(t) - \langle s(t) \rangle]/\langle s(t) \rangle$, $s(t)$ being the data acquired by the photodetector for the three selected wavelengths, and $\langle s(t) \rangle$ the time average value of $s(t)$. (e,f,g) Power Fourier spectrum of the signals shown in panels (b,c,d) respectively. Spectral signals of Bragg and Fresnel signals alone are much weaker than the one resulting from interference between the Fresnel and Bragg reflected beams.}
\end{figure}

As discussed in a previous work \cite{rafayelyan_prl_2016} dealing with constant pitch cholesteric structure, the experimental demonstration of the geometric phase shaping capabilities of Bragg-Berry mirrors can be retrieved by implementing a dynamical phase experiment on uniform constant-pitch cholesterics characterized by $\psi_{\rm surf}(x,y) = \psi_0$. Here we extend this approach to the case of the fabricated uniform gradient-pitch cholesteric sample.

The set-up is sketched in Fig.~\ref{fig:dynamic}(a). The main idea is that the $\psi_0$-dependent phase of the Bragg reflected light, see Fig.~\ref{fig:BBM}(b), can be probed by rotating the sample around the helix axis of the cholesteric at frequency $f_0$. Indeed, according to the relationship $\Delta\phi_R= -2\chi\psi_0$ established in the static case, such a rotational motion imparts to the reflected beam a time-dependent reflected phase with frequency $2f_0$. In contrast, the Fresnel reflection at air/glass interfaces of the rotating sample (glass substrates are not anti-reflection coated) do not contain a time-dependent phase as is the case for standard mirrors.

In turn, following the protocol described in \cite{rafayelyan_prl_2016}, the experiment consists in the detection of the time-varying signal that results from the collinear interference between the Bragg and Fresnel contributions to the total reflected field. The results are shown as solid lines in Fig.~\ref{fig:dynamic}(b), \ref{fig:dynamic}(c) and \ref{fig:dynamic}(d) for three wavelengths falling in the photonic bandgap, namely, $\lambda = 488$, 532 and 633~nm [Fig.~\ref{fig:sample}(c)]. We notice that the Bragg and Fresnel contributions have orthogonal circular polarization states and different powers. Therefore it is necessary to use a quarter-wave plate and a linear polarizer whose relative orientation is chosen to maximize the visibility of the  interference signal recorded by the photodetector, as shown in Fig.~\ref{fig:dynamic}(a).

The power Fourier spectra of the latter signals reveal a peak frequency at $2f_0$, as well as higher-order harmonics at $2 n f_0$ with $n>1$ integer as expected from a non-sinusoidal periodic signal. We also checked that both the Bragg or Fresnel contributions taken separately do not participate to the observed interferential beating, as respectively shown by dashed and dotted lines in Fig.~\ref{fig:dynamic}(b)--(g). Still, residual dynamics appear depending on the probed location of the sample, which we attribute to experimental imperfections of the cholesteric ordering, non-ideal assembling of the substrates and optical adjustments. However, none of these imperfections questions the revealed geometric phase imparted to the helicity-preserved reflected light from gradient-pitch cholesterics.

Although the temporal beatings of above experiments do not allow to discriminate in which direction the sample rotates, we note that the Bragg reflected beam experiences a frequency shift $\pm 2f_0$ where the $\pm$ sign depends on the sense of rotation of the sample and the cholesteric handedness. From a mechanical point of view, the latter blue ($+$ sign) or red ($+$ sign) frequency shift is associated with a nonzero optical radiation torque that results from optical spin angular momentum flipping. The latter optical torque is respectively anti-parallel or parallel to the angular momentum vector of the rotating sample. In contrast, the Fresnel reflection at air/glass interfaces of the sample preserves the spin angular momentum as is the case for standard mirrors, hence it is neither associated with a contribution to the total optical torque exerted on the sample nor to a rotational frequency shift.

\section{Non-uniform gradient-pitch Bragg mirrors}
\label{section:nonuniform}

\begin{figure}[b!]
\centering\includegraphics[width=1\columnwidth] {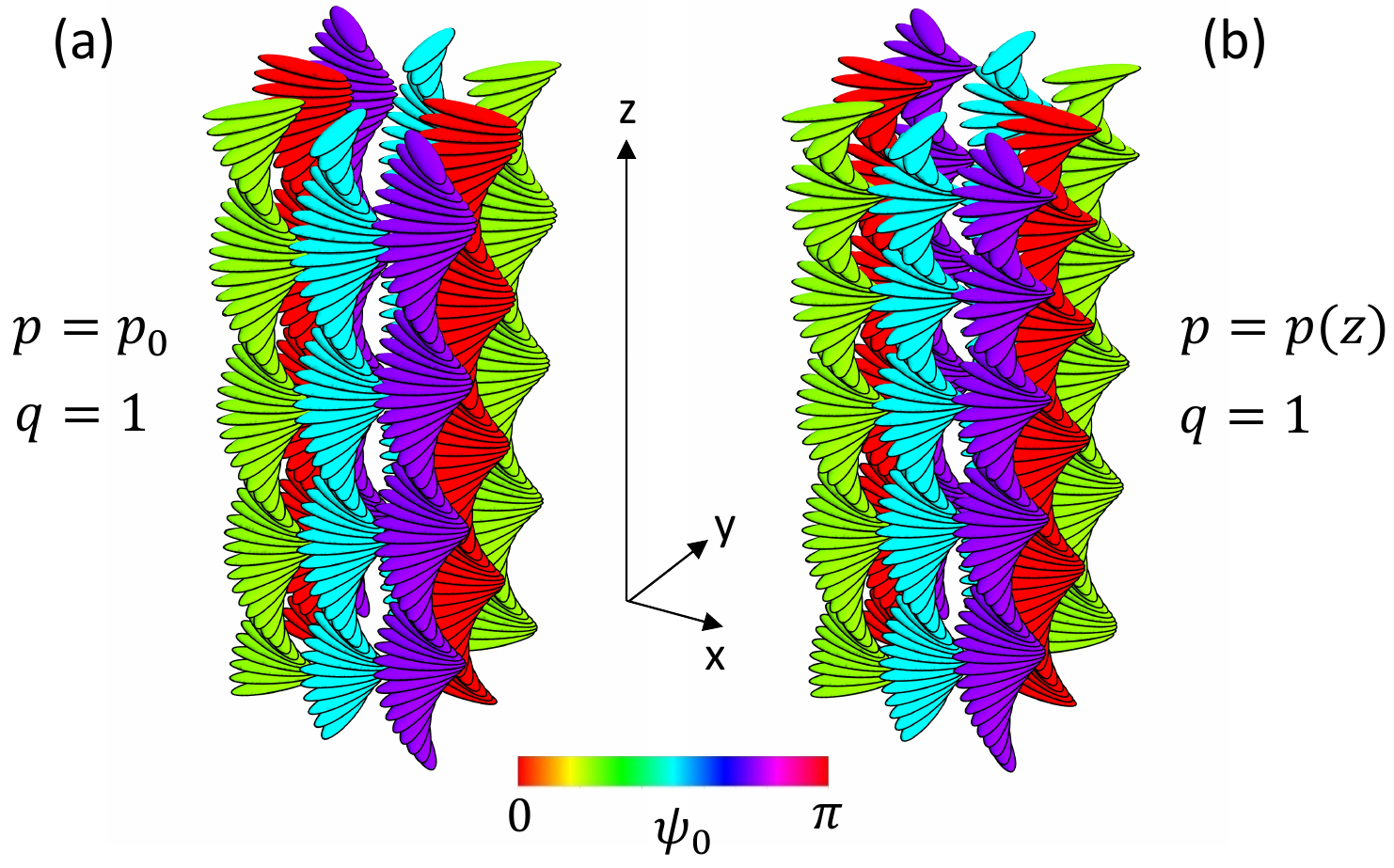}
\caption{
\label{fig:BBsketch_q=1}
Illustration of the three-dimensional molecular ordering following  Eq.~(\ref{eq:psi3D_p(z)_qplate}) in the case $q=1$ for constant-pitch (a) and gradient-pitch (b) cholesterics.}
\end{figure}

After the theoretical and experimental demonstrations of the geometric phase shaping capabilities of gradient-pitch Bragg-Berry mirrors with uniform surface alignment, here we address the case of ultra-broadband geometric phase beam shaping using space-variant orientational boundary conditions, $\psi_{\rm surf}(x,y)$. Without lack of generality we restrict our demonstration to the case of optical vortex generation in order to ease the comparison with previous results on patterned cholesterics with constant pitch \cite{rafayelyan_ol_2016, kobashi_prl_2016}. Namely, we consider a cholesteric slab that combines gradient-pitch and so-called q-plate features \cite{marrucci_prl_2006}, which is ideally described by
\be
\label{eq:psi3D_p(z)_qplate}
\psi_{\rm 3D}(\phi,z) = 2\pi\chi z/p(z) + q\phi\,,
\ee
where the space-variant pitch $p(z)$ is given by Eq.~(\ref{eq:p(z)}) and $\phi$ is the polar angle in the $(x,y)$ plane. We choose $q=1$, which produces Bragg-reflected optical vortex beams with topological charge $\pm2$ depending on the sign of $\chi$. Such a sample was prepared following the protocol presented in section \ref{subsection:uniform-preparation} using photoalignment layers that impose a two-dimensional radial distribution $\psi_{\rm surf}(\phi) = \phi$. This situation thus corresponds to the generalization of the design used in \cite{rafayelyan_ol_2016} by performing the longitudinal structural upgrade $p_0 \to p(z)$. This is illustrated in Fig.~\ref{fig:BBsketch_q=1} where corresponding three-dimensional molecular ordering are shown.

\subsection{White light optical vortex generation}

\begin{figure}[b!]
\centering\includegraphics[width=1\columnwidth]{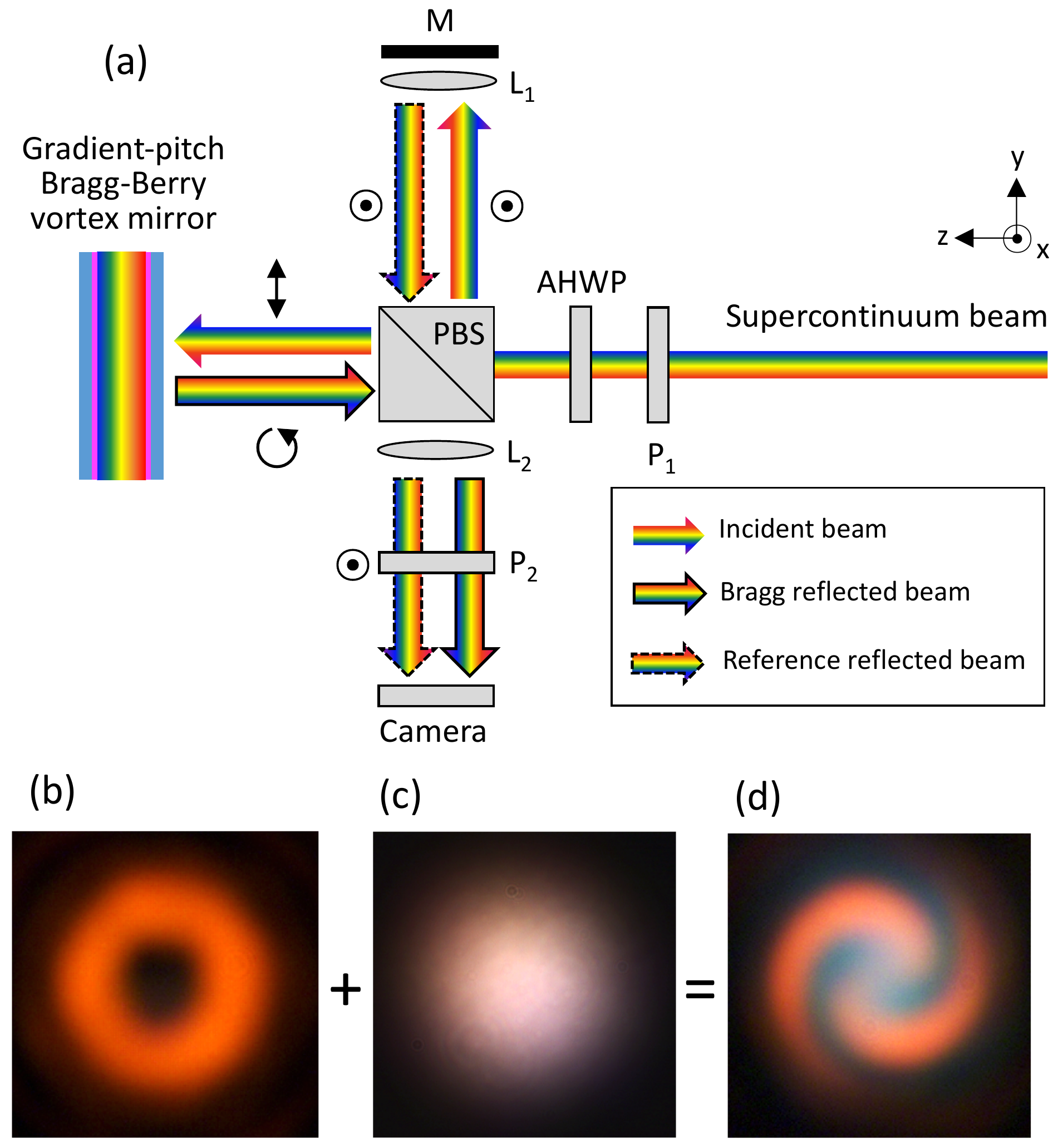}
\caption{
\label{fig:vortex_setup+exp}
(a) Michelson interferometer experimental set-up used to demonstrate the generation of a white light optical vortex beam from a gradient-pitch Bragg-Berry vortex mirror described by Eq.~(\ref{eq:psi3D_p(z)_qplate}), with $q=1$. P$_{1,2}$: linear polarizers;  AHWP: achromatic half-wave plate in the visible domain (from Thorlabs); PBS: polarizing beam splitter; L$_{1,2}$: lenses with focal length $f_1=500$mm and $f_2=300$mm focal length, respectively; M: mirror. Symbols ($\odot, \updownarrow,\circlearrowleft$) refer to the polarization state of light. (b) Far-field intensity profile of the generated white light vortex beam. (c) Reference beam intensity profile. (d) Intensity pattern resulting from the superposition of the vortex and the reference beams.}
\end{figure}

Ultra-broadband optical vortex generation experiment is done using the set-up shown in Fig.~\ref{fig:vortex_setup+exp}(a). It corresponds to a Michelson interferometer scheme where the gradient-pitch Bragg-Berry mirror (GPBBM) with $q=1$ stands as one of the two mirrors of the interferometer. A linearly polarized supercontinuum Gaussian beam with controlled polarization orientation is prepared using a linear polarizer ($P_1$) and an achromatic half-wave plate (AHWP). This allows to control the power ratio of the two arms of the interferometer by using a polarizing beam splitter (PBS). By doing so, the air/glass Fresnel contribution of the beam reflected off the sample and the reference beam reflected off the standard mirror (M) have orthogonal polarization states. Therefore, post-selection of the Bragg-reflected beam is realized by using a linear polarizer oriented along the $x$-axis (P$_2$). Here, a non-trivial point is to benefit from the natural imperfection of the PBS. Indeed, for an ideal PBS the $x$-polarized light reflected off M would return back to the laser source. However, in practice, PBS does not redirect a small fraction of that field, which enables the selective formation of an interference pattern with the $y$-polarized contribution of the Bragg light reflected off the cholesteric sample.

\begin{figure}[b!]
\centering\includegraphics[width=1\columnwidth]{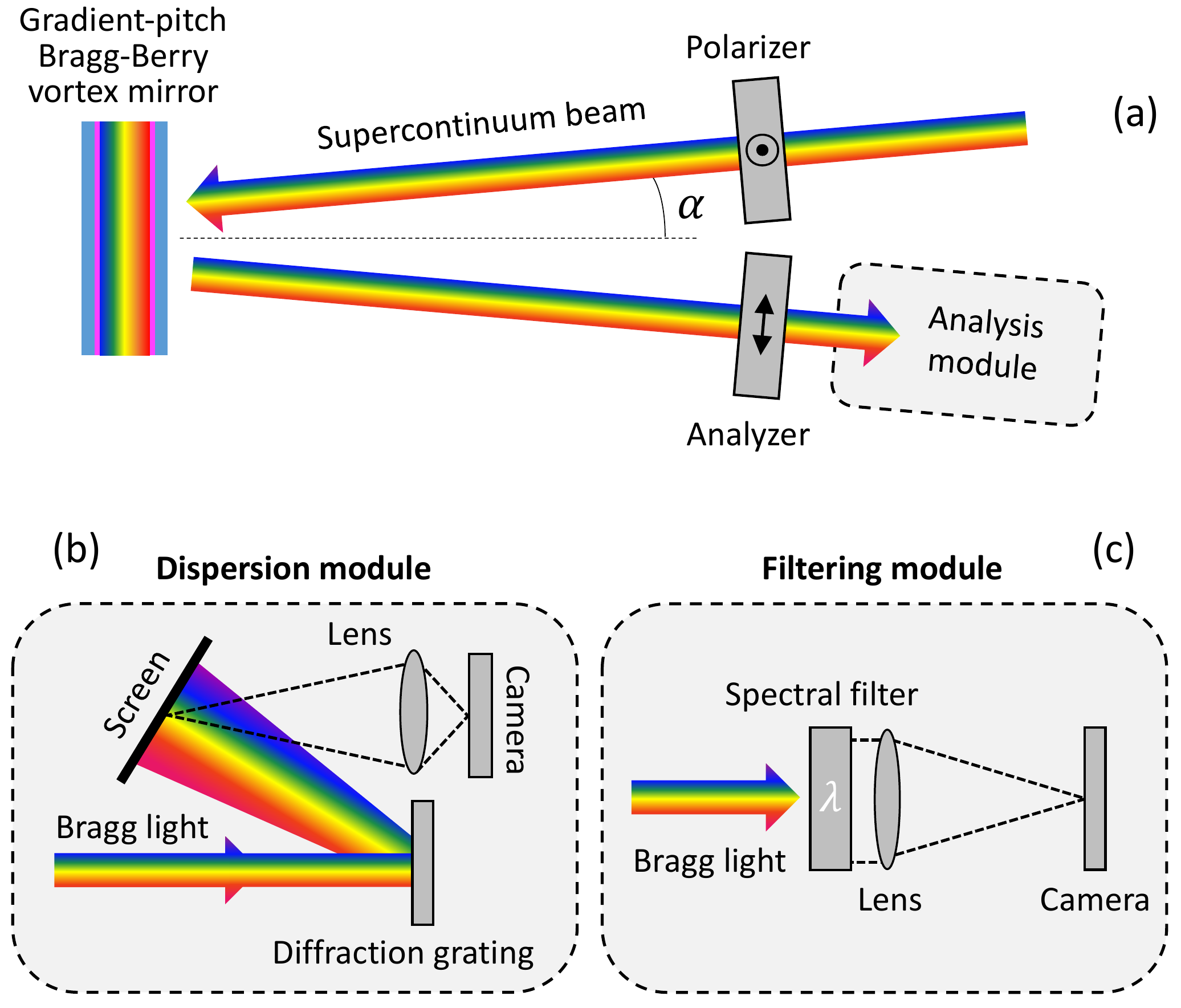}
\caption{
\label{fig:vortex_setup_spectral}
(a) Sketch of experimental set-up for spectral analysis. Slightly oblique incidence ($\alpha \sim 1^\circ$) allows to readily analyze the reflected light. The Bragg-reflected light is selected by observing the sample between crossed linear polarizers, taking rid off the Fresnel reflection from the interface between air and the glass substrate. The analysis module refers to two distinct options detailed in panels (b) and (c). (b) Dispersion module: Bragg light is reflected off a blazed wavelength reflective diffraction grating (from Thorlabs, model GR13-0605) and projected on a black screen. The screen is then re-imaged on a camera. (c) Spectral filtering module: we use a set of seven interference filters centered on $\lambda=400$ to 700 nm by step of 50~nm, with 10~nm full-width half-maximum transmission spectrum. The camera placed in the focal plane of the lens records the far-field of the generated vortex beam.}
\end{figure}

The co-axial interference pattern is acquired by a camera (Cam) placed in the focal plane of a lens (L$_2$). This allows imaging the far-field of the vortex beam generated by the sample, which exhibits a doughnut shape intensity profile as expected, see Fig.~\ref{fig:vortex_setup+exp}(b). On the other hand, the curvature of the reference beam in the plane of the camera, see Fig.~\ref{fig:vortex_setup+exp}(c), is adjusted by placing a lens (L$_1$) nearby M, which allows to control the number of fringes in the field of view. The resulting interference pattern is a two-arm spiral unveiling the generation of an optical phase singularity with topological charge two Fig.~\ref{fig:vortex_setup+exp}(d). We note the unavoidable presence of a whitish blurring on the latter pattern, which is due to the fact that the visibility of the spatial modulation of the intensity cannot be optimized for all the wavelengths at the same time since Bragg reflection alters the spectrum. These observations thus complete the experimental validation.

\subsection{Optical vortex generation: spectral behavior}

More precise analysis of the generated white light vortex beam by the gradient-pitch Bragg-Berry vortex mirror is performed using the experimental set-up depicted in Fig.~\ref{fig:vortex_setup_spectral}. It consists to analyze spectrally the Bragg-reflected field off the sample. This is done by illuminating the vortex mirror at small oblique incidence ($\alpha \sim 1^\circ$) by a linearly polarized supercontinuum Gaussian beam and selecting the Bragg-reflected field only by placing an output linear polarizer whose direction is orthogonal to that of the incident one, as shown in Fig.~\ref{fig:vortex_setup_spectral}(a). The Bragg-reflected light beam is then analyzed in two different and complementary ways. On the one hand, its spectral content is revealed by imaging a black screen on which is projected the spectrally dispersed beam by using a reflective diffraction grating, see Fig.~\ref{fig:vortex_setup_spectral}(b). On the other hand, the far field intensity profile for a discrete set of wavelengths are recorded by a camera at the focal plane of a lens placed after an interference filter, see Fig.~\ref{fig:vortex_setup_spectral}(c).

\begin{figure}[b!]
\centering\includegraphics[width=1\columnwidth]{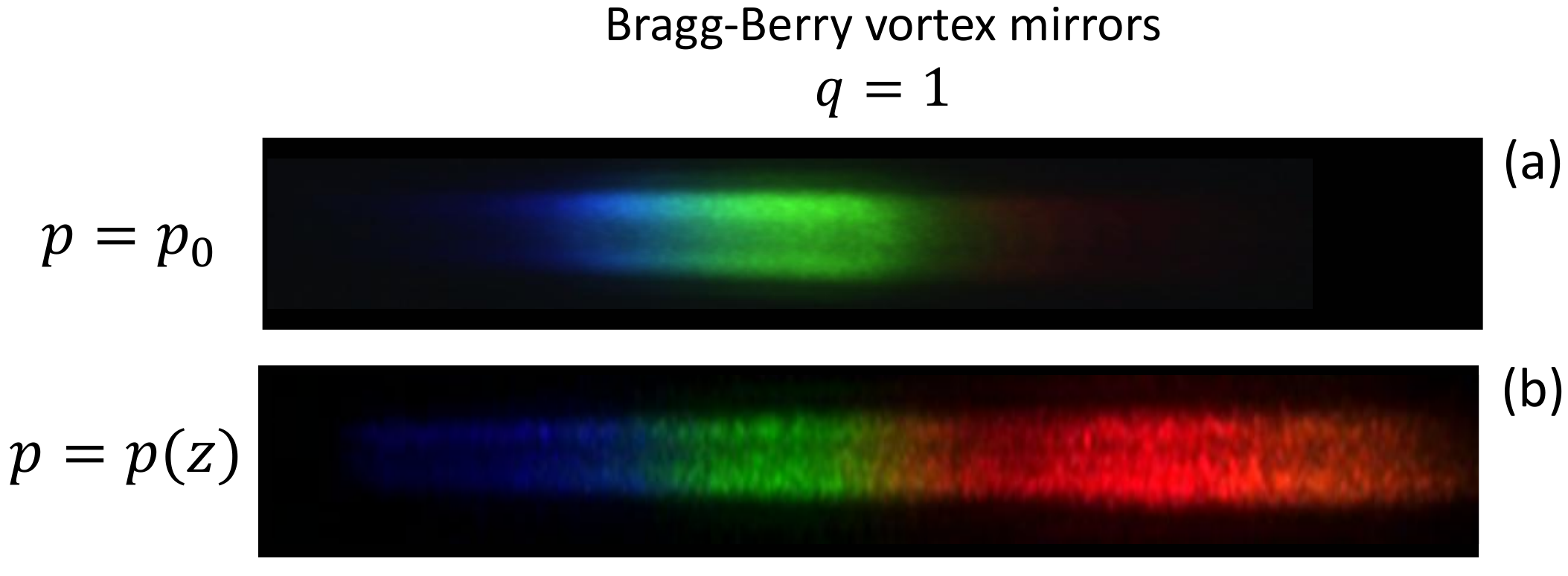}
\caption{
\label{fig:vortex_analysis_spectral}
Experimental spectra of the Bragg-reflected beam off a Bragg-Berry vortex mirror with $q=1$ for constant-pitch (a) and gradient-pitch (b).}
\end{figure}

In order to appreciate the enhanced polychromatic geometric phase shaping of the gradient-pitch Bragg-Berry vortex mirror, the experiments are compared with the case of a constant-pitch Bragg-Berry vortex mirror with the identical charge $q=1$, which was prepared using the cholesteric liquid crystal SLC79 (from BEAM Co) possessing a $\sim70$~nm-width photonic bandgap centered on 530~nm wavelength. Dispersion results are shown in Fig.~\ref{fig:vortex_analysis_spectral} where panel (a) refers to constant-pitch and panel (b) to gradient-pitch. As expected from previous observation of circular Bragg reflection with several hundreds of nanometers bandwith (Fig.~\ref{fig:sample}), these observations confirm that transverse space-variant orientational structure of the cholesteric sample does not alter the bandwidth enhancement.

\begin{figure}[t!]
\centering\includegraphics[width=1\columnwidth]{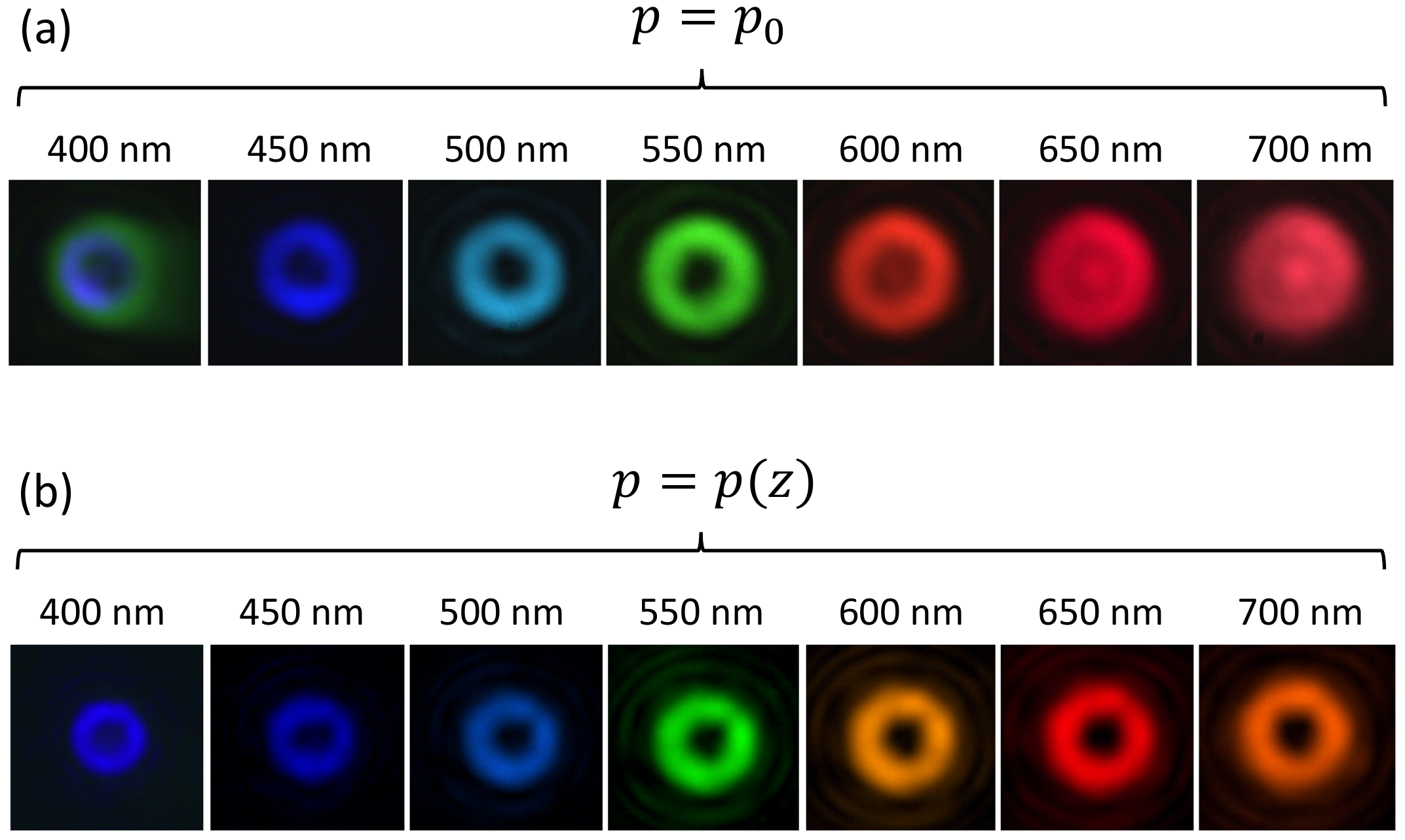}
\caption{
\label{fig:vortex_analysis_farfield}
Experimental set of spectral components for the far-field intensity profiles of the generated vortex beam by a Bragg-Berry vortex mirror with $q=1$ for constant-pitch (a) and gradient-pitch (b).}
\end{figure}

Similarly, the comparative study of the far field vortex beam intensity profiles is displayed in Fig.~\ref{fig:vortex_analysis_farfield}. Note that in the case of constant pitch, image recording requires longer acquisition time outside the photonic bandgap, hence revealing signal-to-noise issues altering the ideally expected doughnut intensity pattern whatever the wavelength. The good quality of the vortex beam profile over the 400-700~nm range thus recalls the ultra-broadband features of the gradient-pitch structure. Also, wavelength-dependent diameter of the doughnut beam is associated with the spectral dependence of the waist of the supercontinuum source.

\subsection{Optical vortex generation: oblique incidence}

The dependence of the Bragg-reflected field from a gradient-pitch Bragg-Berry vortex mirror on the angle of incidence is also explored experimentally and compared to numerical expectations. We note that only the simplest situation of a uniform cholesteric with a constant pitch illuminated at normal incidence, which is presented in Fig.~\ref{fig:BBM}, finds an analytical solution for its reflection and transmission fields according to the Berreman method \cite{berreman_prl_70}. In order to deal with the case of oblique incidence, a not too cumbersome approximate analytical approach called coupled-mode theory has been developed~\cite{yeh_book_10, scharf_book_07}. However, when exact solution is sought, one has to handle the problem within a numerical approach, for instance by using the Berreman 4$\times$4 matrix formalism~\cite{berreman_josa_72}, the Ambartsumian’s layer addition modified method~\cite{gevorgyan_pre_07} or the finite-difference time-domain method (FDTD)~\cite{sullivan_book_13}. In present study, the 4$\times$4 Berreman approach is used.

\begin{figure}[t!]
\centering\includegraphics[width=1\columnwidth]{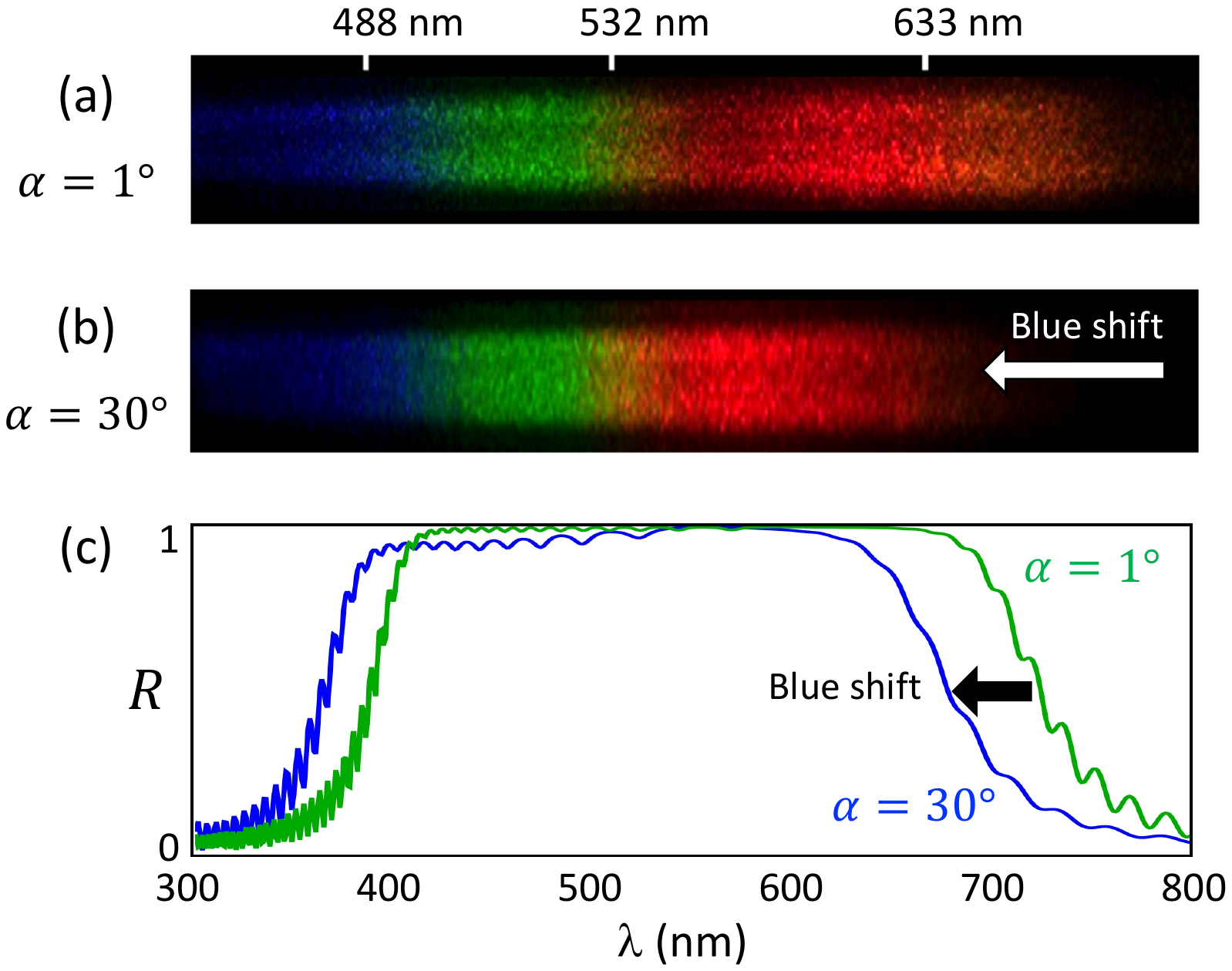}
\caption{
\label{fig:vortex_angle_spectral}
Experimental spectra of the Bragg-reflected beam off a gradient-pitch Bragg-Berry vortex mirror with $q=1$ following the set-up of Figs.~\ref{fig:vortex_setup_spectral}(a) and \ref{fig:vortex_setup_spectral}(b) for $\alpha = 1^\circ$ (a) and $\alpha = 30^\circ$ (b). The three ticks at $\lambda = 488$, 532 and 633~nm, which have been evaluated using interference filters, provide with a nonlinear ruler in order to appreciate quantitatively the frequency shift of the spectrum. (c) Simulated reflectance spectra for external incidence angle $\alpha = 1^\circ$ (green curve) and $\alpha = 30^\circ$ (blue curve) for a uniform gradient-pitch cholesteric with $\psi_0=0$. Numerical parameters are the same as in Fig.~\ref{fig:BBMgrad}.
}
\end{figure}

Firstly, following the set-up of Figs.~\ref{fig:vortex_setup_spectral}(a) and \ref{fig:vortex_setup_spectral}(b), we compare the reflectance spectra for almost normal incidence ($\alpha = 1^\circ$) and oblique incidence ($\alpha = 30^\circ$). The results are shown in Figs.~\ref{fig:vortex_angle_spectral}(a) and \ref{fig:vortex_angle_spectral}(b). A substantial blue shift of the spectrum of several tens of nanometers is observed, which recalls the usual behavior of cholesterics under oblique incidence. This is qualitatively compared with simulations Fig.~\ref{fig:vortex_angle_spectral}(c) performed in the case of a uniform gradient-pitch cholesteric with $\psi_0=0$.\\

\begin{figure}[t!]
\centering\includegraphics[width=1\columnwidth]{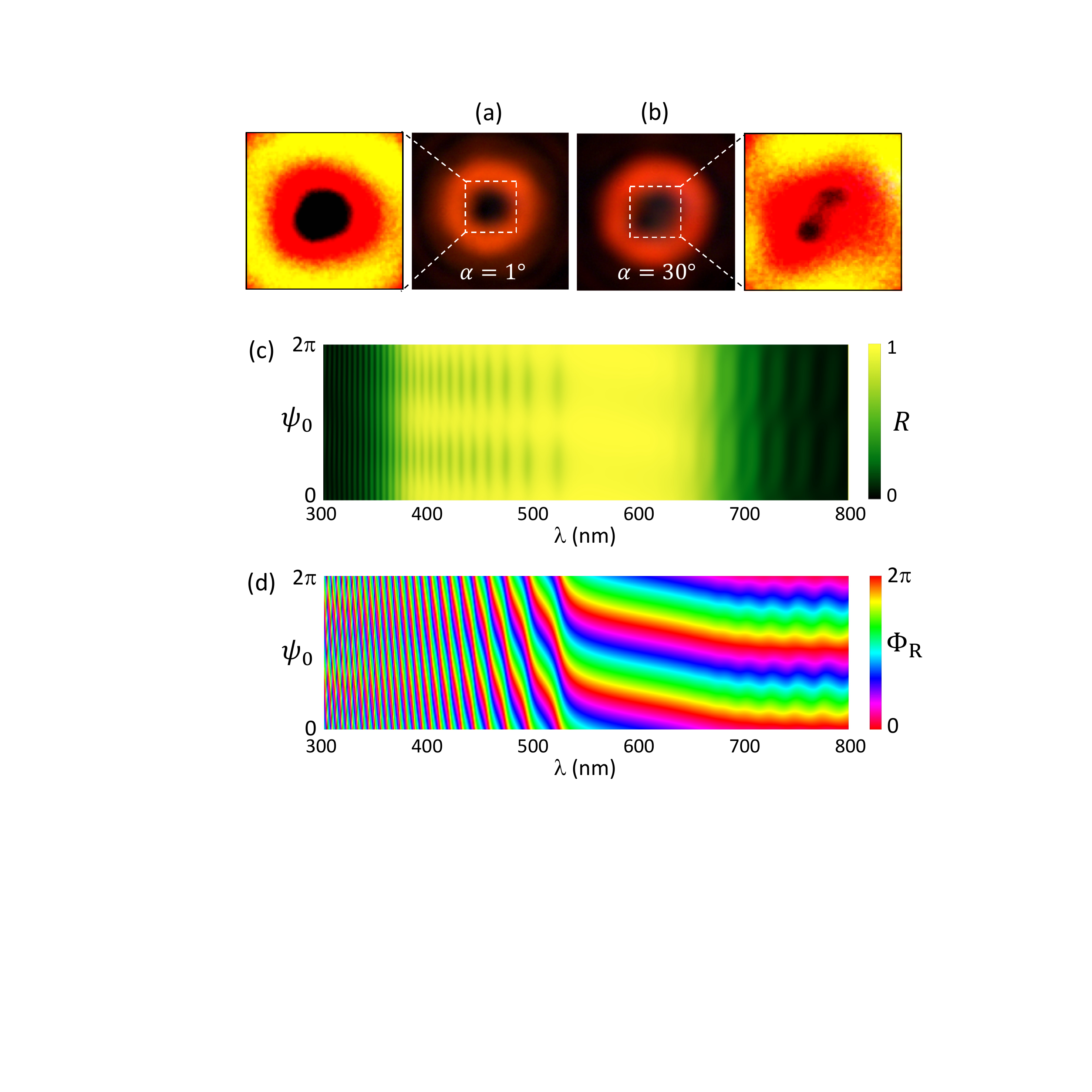}
\caption{
\label{fig:vortex_angle_farfield}
Experimental far field intensity profiles of the Bragg-reflected beam off a gradient-pitch Bragg-Berry vortex mirror with $q=1$ following the set-up of Figs.~\ref{fig:vortex_setup_spectral}(a) and \ref{fig:vortex_setup_spectral}(c) for $\alpha = 1^\circ$ (a) and $\alpha = 30^\circ$ (b). An enlargement of the central part of panels (a) and (b) are shown, where the luminance and contrast have been adapted to emphasize the presence of the splitting of the double-charge phase singularity into two unit charge phase singularities in the case $\alpha = 30^\circ$. Simulated reflectance (d) and reflected phase (e) spectra for external incidence angle $\alpha = 30^\circ$ for a uniform gradient-pitch cholesteric with $0 < \psi_0 < 2\pi$. Numerical parameters are the same as in Fig.~\ref{fig:BBMgrad}.
}
\end{figure}

Then, following the set-up of Figs.~\ref{fig:vortex_setup_spectral}(a) and \ref{fig:vortex_setup_spectral}(c), we compare the far field intensity profiles of the generated vortex beams for $\alpha = 1^\circ$ and $\alpha = 30^\circ$, see Figs.~\ref{fig:vortex_angle_farfield}(a) and \ref{fig:vortex_angle_farfield}(b). Qualitatively, the doughnut-shaped characteristic of the generated vortex beam is preserved as the incidence angle increases. In addition, a closer look at the central part unveils a splitting of the optical phase singularity with topological charge two into two singularities with unit charge, see enlargement panel of Fig.~\ref{fig:vortex_angle_farfield}(b). Such high-charge splitting is a well-known practical feature, especially in the presence of a smooth coherent background \cite{basistiy_oc_93}, that affects even high-quality high-charge vortex beams \cite{ricci_oe_2012}.

However, we notice that the vortex splitting is much more pronounced at oblique incidence than at normal incidence in which case we hardly evidence it with the used imaging device, as shown in the enlargement panel of Fig.~\ref{fig:vortex_angle_farfield}(a). This is explained recalling that tilted illumination breaks the axisymmetry \cite{rafayelyan_ol_2016} and the symmetry breaking is all the more pronounced as the angle of incidence is high. From the theoretical point of view, the latter features can be grasped from simulations of the reflectance and reflected phase of the Bragg-reflected light on a uniform gradient-pitch Bragg-Berry mirror as a function of $\psi_0$. The calculated spectra are shown in Figs.~\ref{fig:vortex_angle_farfield}(c) and \ref{fig:vortex_angle_farfield}(d), respectively. The data thus confirm that the geometric phase shaping is unaltered by the oblique incidence while the reflectance is substantially modulated as a function of  $\psi_0$. Still, there might also have other contributions to the observed high-charge splitting that are not taken into account in the latter simulations, for instance the actual three-dimensional modulation of the chiral properties of the sample.

\section{Conclusion}
\label{section:conclusion}

In this work we showed that chiral nematic liquid crystal with helical ordering that varies in three dimensions enables the realization of a ultra-broadband geometric phase reflective flat optical elements, namely gradient-pitch Bragg-Berry mirrors. The selective spatial phase structuring imparted to the Bragg-reflected beam off the device is determined by the surface orientational boundary conditions of the liquid crystal slab. On the other hand, the polychromatic properties of the device is ensured by the presence of a gradient of the pitch of the helical bulk molecular ordering. Experimental demonstration has been performed by fabricating an optical vortex generator operating in the full visible domain. Topological, spectral and angular optical properties have been experimentally determined and discussed in the framework of numerical simulations. These results extend the use of previously introduced Bragg-Berry mirrors to very large spectral bandwidths that can be adjusted by appropriate choice of the material properties.

Of course, although present results are restricted to the polychromatic management of the orbital angular momentum of light, they could be extended to any phase shaping by appropriate surface patterning of alignment layers using state-of-the-art techniques \cite{kim_optica_2015, chen_pr_2015}.  Nevertheless, it should be recalled that light scattering remains a challenging issue that need further work. Indeed, chiral gradients in the three spatial dimensions lead in practice to substantial scattering losses. In the present case, the specular Bragg reflectance for wavelengths falling in the photonic bandgap is typically four times smaller than the ideal value.

Also, note that other three-dimensional chiral structures that are not nematic have been previously discussed in the context of Bragg-Berry mirrors, see for instance Ref.~\cite{yoshida_lc_2016} where the case of space-variant blue phases samples has been discussed. Finally, we would like to mention that we became aware of a similar study made using photo-induced gradient-pitch cholesterics published very recently \cite{kobashi_mclc_2017} after this work has been completed \cite{rafayelyan_thesis}. In practice, since present study has been made using glassy gradient-pitch cholesterics, it offers an interesting complementary approach. Also, in contrast to Ref.~\cite{kobashi_mclc_2017}, we notice that present study provides with  behavior of the polychromatic beam shaping versus the incidence angle and in the far field, thus bringing a thorough overview of the optical properties of gradient-pitch Bragg-Berry optical elements.

\section*{Acknowledgment} This study has been carried out with financial support from the French National Research Agency (project ANR-15-CE30-0018).


\begin{thebibliography}{}
\bibitem{remark_qualitative}
Indeed, a correct simulation of the phenomenon should account for the azimuthal
  orientational gradients of the liquid crystal orientation in the bulk of a
  transversely space-variant cholesteric slab with uniform pitch.
\end{thebibliography}

\end{document}